\documentclass[12pt]{article}
\makeatletter \@addtoreset{equation}{section} \makeatother
\renewcommand{\theequation}{\thesection.\arabic{equation}}
\newcommand{\ba}{\begin{array}}
\newcommand{\ea}{\end{array}}
\newcommand{\beq}{\begin{equation}}
\newcommand{\eeq}{\end{equation}}
\newcommand{\bea}{\begin{eqnarray}}
\newcommand{\eea}{\end{eqnarray}}

\def\bce{\begin{center}}
\def\ece{\end{center}}

\def\nonu{\nonumber}

\def\be{\beta}

\def\eps6{{\displaystyle \mathop{\epsilon}^{6}}{}}
\def\g6{{\displaystyle \mathop{g}^{6}}{}}

\def\nab6{{\displaystyle \mathop{\nabla}^{6}}{}}
\def\to{\rightarrow}

\def\0{{\sst{(0)}}}
\def\1{{\sst{(1)}}}
\def\2{{\sst{(2)}}}
\def\3{{\sst{(3)}}}
\def\4{{\sst{(4)}}}
\def\5{{\sst{(5)}}}
\def\6{{\sst{(6)}}}
\def\7{{\sst{(7)}}}
\def\8{{\sst{(8)}}}

\def\ba{\begin{array}}
\def\ea{\end{array}}
\def\beq{\begin{equation}}
\def\eeq{\end{equation}}
\def\be{\begin{equation}}
\def\ee{\end{equation}}

\def\to{\rightarrow}

\def\eps{\epsilon}

\def\ba{\begin{array}}
\def\ea{\end{array}}
\def\beq{\begin{equation}}
\def\eeq{\end{equation}}
\def\be{\begin{equation}}
\def\ee{\end{equation}}

\def\to{\rightarrow}

\def\eps{\epsilon}

\def\eps6{{\displaystyle \mathop{\epsilon}^{6}}{}}

\def\nab6{{\displaystyle \mathop{\nabla}^{6}}{}}

\newcommand{\bean}{\begin{eqnarray*}}
\newcommand{\eean}{\end{eqnarray*}}

\begin{document}
\thispagestyle{empty} \addtocounter{page}{-1}
   \begin{flushright}
PUPT-2380 \\
\end{flushright}

\vspace*{1.3cm}
  
\centerline{ \Large \bf   
The Large $N$ 't Hooft Limit of 
Coset Minimal Models }
\vspace*{1.5cm}
\centerline{{\bf Changhyun Ahn  
\footnote{On leave from the Department of Physics, Kyungpook National University, Taegu
  702-701, Korea and 
address until Aug. 31, 2011:
Department of Physics, Princeton University, Jadwin Hall, 
Princeton, NJ 08544, USA} }}
\vspace*{1.0cm} 
\centerline{\it  
Department of Physics, Princeton University, Jadwin Hall, 
Princeton, NJ 08544, USA}
\centerline{\it  
Department of Physics, Kyungpook National University, Taegu
702-701, Korea} 
\vspace*{0.8cm} 
\centerline{\tt ahn@knu.ac.kr} 
\vskip2cm

\centerline{\bf Abstract}
\vspace*{0.5cm}

Recently, Gaberdiel and Gopakumar proposed that 
the two-dimensional $WA_{N-1}$ minimal model conformal field theory in the
large $N$ 't Hooft limit is dual to the higher spin theories on the 
three-dimensional AdS space with two complex scalars. 
In this paper, we examine this proposal
for the $WD_{\frac{N}{2}}$ and $WB_{\frac{N-1}{2}}$ minimal models
initiated by Fateev and Lukyanov in 1988.  
By analyzing the renormalization group flows on these models,
we find that
the gravity duals in AdS space are higher spin theories coupled to two
equally massive real scalar fields.  
We also describe the large $N$ 't Hooft limit for the minimal model of the second
parafermion theory.

\baselineskip=18pt
\newpage
\renewcommand{\theequation}
{\arabic{section}\mbox{.}\arabic{equation}}


\section{Introduction}

By looking at the well understood family of two-dimensional conformal
field theories with an appropriate large $N$ limit, Gaberdiel and
Gopakumar \cite{GG,Gtalks} have been 
using the AdS/CFT correspondence to look for three-dimensional
classical gravity theories. 
They consider a particular $A_{N-1}$ WZW
coset minimal model \cite{GKO85,GKO86,BBSS,BBSS1} 
which has a higher spin $W A_{N-1}(\equiv W_N)$ symmetry
generated by currents of spins $2, 3, \cdots, N$ \cite{FL}. See also \cite{BS} 
for a detailed description of $W$-symmetry in conformal field theory.
Their large $N$ 't Hooft limit is defined as
\bea
N, k \rightarrow \infty, \qquad \lambda \equiv \frac{N}{k+N} \qquad \mbox{fixed},
\label{limit}
\eea
where 't Hooft coupling $\lambda$ is a function of $N$ and $k(=1, 2, \cdots)$ and
runs from zero to $1$. The $k$ is the level of the WZW current
algebra. The central charge is given by $c_N(\lambda) \simeq N (1-\lambda^2)$
under (\ref{limit}). 
The bulk theory they found is a Vasiliev type higher spin theory \cite{PV,PV1,Vasiliev} in
three-dimensional AdS space coupled with 
two complex (equally massive) scalar
fields where the mass of the fields is given by $M^2 = -(1-\lambda^2)$ which lies
between $-1$ and zero. The above two complex scalars are quantized
with opposite (conformally invariant) boundary conditions. 
Therefore,  their conformal dimensions are $h_{+}= \frac{1}{2}(1+\lambda)$ and 
$h_{-}=\frac{1}{2}(1-\lambda)$.
The check for this duality is based on two aspects. 1) The 
two partition functions are found to match. The total partition function in
the bulk consists
of the sum of the contributions from 
both higher spin fields and the two complex scalar fields. 
It is quite nontrivial to find the conformal field theory partition function
from the character formula within the large $(N,k)$ 't Hooft limit.
2) The renormalization group (RG) flow patterns are
coincident with each other. The RG flow for large $N >2$ in the boundary
theory is assumed to be obtainable by following the RG flow for $N=2$ and the AdS/CFT
correspondence is used in the bulk theory by interpreting the RG flow as a change
of  boundary conditions on one of the fields.

It is natural to ask whether there exist any higher spin $AdS_3$ gravity duals to
other types of unitary coset minimal models.
Some time ago, Lukyanov and Fateev \cite{LF} classified other types of
(extended) 
$W$-symmetry algebras: $WD_n$ symmetry algebras generated by 
currents of spins 
\bea
WD_n:2, 4, \cdots, 2(n-1), \,\,\, \mbox{and} \,\,\, n 
\label{wd}
\eea
and
$WB_n$ symmetry algebra generated by 
currents of spins 
\bea
WB_n:2, 4, \cdots, 2n, \,\,\, \mbox{and} \,\,\, (n+\frac{1}{2}).
\label{wb}
\eea  
The conformal dimension and the spin are linear combinations of 
the holomorphic conformal dimension and its antiholomorphic counterpart.
More precisely, the above currents (left component) have  spins which are
holomorphic conformal dimensions. Of course, their counterparts (right component) 
have spins which are opposite to its antiholomorphic
conformal dimensions.  
Sometimes the latter algebra is denoted as $WB(0,n)$ with a Lie superalgebra
$B(0,n)=OSp(1,2n)$ (for example, in \cite{BS}) because the spin contents (\ref{wb})
come from the results of the Drinfeld-Sokolov  reduction to this
superalgebra rather than $B_n$ itself. In this paper, 
by following the procedure of \cite{GG}, 
we describe the coset WZW theories based on the above minimal models
described by (\ref{wd}) and (\ref{wb}). Mainly we focus on their behaviors
under the large $(N,k)$ 't Hooft limit (\ref{limit}) and
once we have found the two-dimensional results from the RG flows, then
we reconsider them in the bulk using the AdS/CFT correspondence. 

In section 2, we consider the diagonal coset minimal $WD_n^{(p)}$
model, where $p$ is a minimal model index.
By reading off  conformal dimensions for primary fields developed
in \cite{LF,LF1,DNS}, we compute conformal dimensions for the 
relevant primary field and the other
nontrivial lowest primary 
field (which has a nontrivial operator product expansion with the
relevant field) in the large 
$N$ 't Hooft limit (\ref{limit}). 
For known fusion rules between these two primaries with explicit
structure constants (or three-point functions), 
we analyze the RG flow(due to the presence of the above
relevant field) between the two fixed points: One fixed point is described by
the $W D_n^{(p)}$ minimal model and the other one 
by the $W D_n^{(p-1)}$ model in which  the minimal model index is shifted by $1$. 
The description of the bulk theory, a higher spin theory coupled to two
equally massive `real' scalar fields, is obtained from the AdS/CFT correspondence. 
We also describe the total partition function in the bulk/boundary.

In section 3, we describe the procedures of section 2 for the case
of the $W B_n^{(p)}$ model \cite{LF,DE}. We only present the main results
without the details. 

In section 4, we summarize what we have presented in this paper and comment
on some future research directions. In particular, we briefly sketch out the
large $(N,k)$ 't Hooft limit  for the second
parafermion theory found in \cite{FZ1985}. Finally, we describe one of the possible
supersymmetric versions of the proposal \cite{GG}.  

Another proposal for large $N$ limits of two dimensional solvable
conformal field theories with their AdS duals is found in \cite{KN}.

\section{ The large $(N, k)$ limit of coset minimal $W D_n^{(p)}$ model}

Let us consider the `diagonal' coset WZW model
characterized by \cite{LF,GKO85}
\bea
\frac{\widehat{SO}(N)_k \oplus \widehat{SO}(N)_1}{\widehat{SO}(N)_{k+1}}.
\label{coset}
\eea
Denoting the spin 1 current fields of the affine Lie algebra $\widehat{SO}(N) 
\oplus \widehat{SO}(N)$ as $E_{(1)}^{ab}(z)$ and $E_{(2)}^{ab}(z)$,
of levels $k$ and $1$, respectively, and the spin 1 current field of the diagonal
affine Lie subalgebra $ \widehat{SO}(N)$, which has level $(k+1)$, as $E'^{ab}(z)$,
we have the relation $E'^{ab}(z) = E_{(1)}^{ab}(z) +E_{(2)}^{ab}(z)$.
The level of the diagonal subalgebra is the sum of the other two levels
because $E_{(1)}^{ab}(z)$ and $E_{(2)}^{ab}(z)$ commute with each other.
The coordinate $z$ is the complex coordinate in two dimensional 
conformal field theory.
The indices $a,b$ take the values $a, b=1, 2, \cdots, N$ 
in the representation of finite-dimensional 
Lie algebra $SO(N)$. These current fields of the WZW model are antisymmetric
in the indices $a, b$ and satisfy the standard operator product
expansion \cite{DMS,BS}.
We introduce a rank $n$ for $D_n=SO(2n)$ with a relation  
\bea
N \equiv 2n.
\label{rank}
\eea

The coset Virasoro generator $\widetilde{T}(z)$ in (\ref{coset}) can be constructed
from the relation $\widetilde{T}(z) = T_{(1)}(z) +T_{(2)}(z)-T'(z)$.
The stress energy tensors can be obtained from the Sugawara
construction \cite{BS}; they are quadratic in the currents.
Of course, $\widetilde{T}(z)$ commutes with the diagonal 
current $E'^{ab}(z)$, which can be shown  by computing
the operator product expansion between them (and similarly
with $T'(z)$).
The central charge of the coset Virasoro algebra is
$\widetilde{c}=c_{(1)} +c_{(2)} -c'$, which can be seen 
by computing the operator product
expansion between $ T_{(1)}(z) +T_{(2)}(z)-T'(z)$  and 
$ T_{(1)}(w) +T_{(2)}(w)$ in which we use the fact that 
$\widetilde{T}(z)$ commutes with 
$T'(w)$. 
The
operator product expansion between $T'(z)$ and $ T_{(1)}(w)
+T_{(2)}(w)(=\widetilde{T}(w) +T'(w))$ is equivalent to  $T'(z) T'(w)$
and then  the above  operator product expansion is  $
T_{(1)}(z)  T_{(1)}(w) +T_{(2)}(z)  T_{(2)}(w)-T'(z) T'(w)$.
The coset central charge is a sum of three parts.  
Then the coset central charge is a function of $p$ (\ref{pcondition})
as follows \cite{LF}:
\bea
c_N(p) & = & \frac{1}{2}N (N-1) \left[ \frac{k}{k+(N-2)} + 
\frac{1}{1+(N-2)} - \frac{k+1}{k+1+(N-2)} \right]
\nonu \\
&=&
\frac{N}{2} \left[ 1- \frac{(N-2)(N-1)}{p(p+1)} \right] \leq \frac{N}{2},
\label{centralcharge}
\eea
where the parameter $p$ is introduced as a function of $N$ and level $k$
indicating the minimal model index
\bea
p\equiv k+N-2 \geq N-1, \qquad k=1,2, \cdots.
\label{pcondition}
\eea 
We used in (\ref{centralcharge}) the fact that the 
dual Coxeter number of $SO(N)$ is
given by 
$h^{\nu} = N-2$ and the dimension of $SO(N)$ is $\mbox{dim} \, SO(N)=\frac{1}{2}N(N-1)$. 
As in the 
$A_{n-1}^{(p)}$ minimal model, the maximum value of the central charge
is the rank of $SO(N)$. 
According to the construction of \cite{LF}, the spin $2$ stress energy
tensor can be written in terms of $n$-component massless scalar fields.
The second order derivatives of these scalar fields have a background
charge $\alpha$. When this background charge satisfies 
$\alpha^2 =\frac{1}{p(p+1)}$, then the central charge 
$c=n-6 \alpha^2 \vec{\rho}^2$ becomes (\ref{centralcharge}) where
the Weyl vector $\vec{\rho}$ will be given later in (\ref{Delta}).
Therefore, the quantum Drinfeld-Sokolov description \cite{BS} for the central
charge is equivalent to the coset description above.
For $N=6$(or $n=3$), 
the conformal field theory of the $WD_3^{(p)}$ minimal model is
discussed in \cite{DNS1}.

Are any critical behaviors of known statistical systems included in this unitary minimal
series (\ref{centralcharge}) and (\ref{pcondition})?
When $p=N-1$(the lowest value of $p$), 
then (\ref{centralcharge}) implies that the central charge is $c=1$, which
describes the particular case of the critical behavior of the 
Ashkin-Teller model \cite{AT}. 
For the next lowest value, $p=N$, the model can be reduced
to 
the $Z_{2N}$ Ising model \cite{FZ1985}.  
When $p \rightarrow \infty$(by taking $k \rightarrow \infty$ with
fixed $N$), the central charge is given by $c=\frac{N}{2}$.
In this case, the symmetry algebra is the Casimir algebra of $\widehat{SO}(N)$ at
level $1$. 
This can be realized in terms of $N$ real independent free
fermions \cite{DMS} (or see the papers \cite{Ahn91,Ahn92} for
similar considerations), each of which 
contributes $\frac{1}{2}$ to the central charge.
The spin 1 current is quadratic in these fermions. 
Note that the contributions
from $c_{(1)}$ and $-c'$  in the first term and third term of 
(\ref{centralcharge}), in this
limit, exactly cancel each other. Then only the second term from 
$c_{(2)}$ remains and leads to $c=\frac{N}{2}$.
The $A_{n-1}^{(p)}$ minimal model is realized by $(N-1)$ free bosons.

The primary operators of the minimal model we are interested in are represented by 
the vertex operators that can be associated with the weight lattice of 
$D_n$ (or $D_{\frac{N}{2}}$ via (\ref{rank})) \cite{LF}. 
The weight vector that appears as the exponent of the vertex operator 
is labelled by two weight lattices denoted by 
$\alpha_{+}$ and $\alpha_{-}$ (which are two Coulomb gas parameters).
The allowed values of this weight vector should satisfy 
the condition for `strongly' degenerate modules with respect to the chiral algebra.
Then the field theory can be constructed from a finite number of
primary fields.
By introducing $\vec{n} =(n_1, n_2, \cdots, n_n) =\sum_{i=1}^{n}
n_i \vec{w}_i$ to represent the $\alpha_{+}$ side and  $\vec{n}' =(n_1',
n_2', \cdots, n_{n}') =\sum_{i=1}^{n}
n_i' \vec{w}_i$ to represent the $\alpha_{-}$ side, 
where $\vec{w}_i$ with $i=1, 2, \cdots, n$ are the
fundamental weights of the algebra $D_n$, and writing the background
charge in terms of the Weyl vector $\vec{\rho}=(1, 1, \cdots, 1)
=\sum_{i=1}^{n} \vec{w}_i$, 
it is known that 
the Coulomb gas formula for the conformal dimension $\Delta_{(\vec{n}|\vec{n}')}^{(p)}$ of the 
primary operator $\Phi_{(\vec{n}|\vec{n}')}^{(p)}$
can be summarized by \cite{LF,LF1,DNS}
\bea
\Delta_{(\vec{n}|\vec{n}')}^{(p)} =\frac{1}{4p(p+1)}
\left[((p+1)\vec{n}-p\vec{n}')^2-\vec{\rho}^2 \right], \qquad 
\vec{\rho}^2 = \frac{1}{3} n (n-1)(2n-1).
\label{Delta}
\eea
The positive integers $n_i$ and $n_i'$ 
are `Dynkin labels'. For the
standard notation of \cite{Slansky}, one needs to subtract the
components of Weyl
vector from this Dynkin label.
In order to compute the conformal dimension 
(\ref{Delta}) for various $(\vec{n}|\vec{n}')$ 
explicitly, the quadratic form matrix (the metric tensor 
\footnote{For
convenience, we present the products of the weights: $\vec{w}_i\cdot\vec{w}_j= 2i$ for
$i\leq j < n-1$, $\vec{w}_i\cdot\vec{w}_{n-1}=\vec{w}_i\cdot\vec{w}_n
= i$ for $i < n-1$, 
$\vec{w}_n\cdot\vec{w}_n=\vec{w}_{n-1}\cdot\vec{w}_{n-1}=
\frac{n}{2}$, and
$\vec{w}_{n-1}\cdot\vec{w}_n=\frac{n-2}{2}$.
\label{quadform} }
for
the weight space) for $D_n$ is used \cite{DNS}. For example, 
the square of the Weyl vector, $\vec{\rho}^2$, appearing in (\ref{Delta}) 
is the sum of the quadratic
form matrix elements. There is a difference in
the overall factor compared to \cite{LF,LF1,DMS}. We also follow the
Dynkin label notation of \cite{DE} instead of using the notation of 
\cite{DNS}. The $\alpha_{+}$ and 
$\alpha_{-}$ are written in terms of a parameter $p$:
$\alpha_{+} =\sqrt{\frac{p+1}{p}}$ and
$\alpha_{-}=-\sqrt{\frac{p}{p+1}}$.
The positive integers $n_i$ and $n_i'$ should satisfy some conditions, i.e.,
each linear combination of $n_i$ and $n_i'$ is bounded by 
the minimal model index $p$.
The primary fields  $\Phi_{(\vec{n}|\vec{n}')}^{(p)}$ with dimensions
given by 
(\ref{Delta}) together with their descendants form a closed operator algebra.  
The character of the module \cite{LF} can be written as 
$\frac{1}{\eta(\tau)^n} \exp \left[2\pi i
\tau(\Delta_{(\vec{n}|\vec{n}')}^{(p)}-\frac{c_N(p)-n}{24}) \right]$ where
$\eta(\tau)$ is the Dedekind function and $\tau$  is the modular
parameter. 
It is easy to check that 
the last term of (\ref{Delta}) cancels the dimension-independent parts of
the character  and the remaining terms of
(\ref{Delta}) contribute to the final character. 
Note that the combination $\frac{1}{24}(c_N(p)-n)$ appears in the quantum
Drinfeld-Sokolov construction \cite{BS} for the conformal dimension. 

Let us consider the neighborhood of the critical point of 
the $D_n^{(p)}$ model (a minimal model of the main series labelled 
by $p$ (\ref{pcondition}) 
associated with a
simple Lie algebra $D_n$ of rank $n$) with $p$ very much larger than $n$. 
The  perturbed action,
with a slightly different notation for the primary field, 
is given by Fateev and Lukyanov \cite{LF}
\bea
 S^{(p)}_g = S_0^{(p)} + g \int d^2 x \,\, \Phi^{(p)}_{(1^n|1,2,1^{n-2})}(x),
\label{modaction}
\eea
where $S_0^{(p)}$ is the action of the conformal field theory 
of the unperturbed $D_n^{(p)}$ model. See also the original papers by
Zamolodchikov \cite{Zam86,Zam87} for the details.
We use a simplified notation for the vectors indicating the
representations of $D_n$ in weight space: $(1^n) \equiv (1,1, \cdots, 1)$
which is a trivial representation of $D_n$ 
and $(1,2,1^{n-2}) \equiv (1,2,1, \cdots, 1)$ which is an adjoint
representation of $D_n$ \footnote{For the $A_{n-1}^{(p)}$ minimal
model considered in \cite{GG}, the perturbed action \cite{LF} is given by $
 S^{(p)}_g = S_0^{(p)} + g \int d^2 x \,\, \Phi^{(p)}_{(1^{n-1}|2,1^{n-3},2)}(x)
$. The `Dynkin label' $(2, 1^{n-3}, 2)$, which is equivalent to $(1,
0^{n-3},1)$ of \cite{Slansky}, represents the adjoint
representation of $A_{n-1}$.}. The number of elements should be equal
to $n$. In the notation of \cite{Slansky}, the former is $(0^n)$ and
the latter is given by $(0, 1, 0^{n-2})$.
Note that in \cite{DNS} more general perturbations are
considered.
There are multiple relevant operators with slightly relevant terms
quadratic in the energy operator. 
In order to obtain the perturbation (\ref{modaction}) from the
description of \cite{DNS}, one should take the appropriate limit. 

One can easily check that the
dimension of the identity operator, $\Delta_{(\vec{n}|\vec{n})}^{(p)}$, for the 
representation with $\vec{n}=\vec{n}'=(1^n)$ vanishes because the numerator
of (\ref{Delta})
is identically zero.
From (\ref{Delta}), one can write the conformal dimension, 
by expanding, recollecting terms, and taking the large $p$ limit, as follows:
\bea
\Delta_{(\vec{n}|\vec{n}')}^{(p)} = \frac{1}{4}(\vec{n}-\vec{n}')^2 +
\frac{1}{4}(\vec{n}^2-\vec{n}'^2) \epsilon +{\cal O}(\epsilon^2),
\qquad \epsilon \equiv \frac{1}{p+1} \simeq \frac{1}{p}.
\label{expansion}
\eea
The matrix of scalar products of the fundamental weights of the Lie
algebra $D_n$ is assumed here(in footnote \ref{quadform}).
There are infinitely many solutions for (\ref{expansion}) 
to possess slightly
relevant fields(which have the conformal dimension $1$ approximately) 
as $p \rightarrow \infty$. However, 
for the choice of the trivial $\alpha_{+}$ side with $(1^n)$, there is a unique
relevant field as in (\ref{modaction}) above 
because the $(2, 2)$ component of the quadratic form
matrix (in footnote \ref{quadform}) 
is equal to $4$ and this provides a constant term $1$ in
(\ref{expansion})
\footnote{More precisely, there exists a unique `slightly' relevant
  field. A relevant field, in general,  has conformal dimension less than
  $1$($\Delta < 1$)
  because the scaling dimension should be less than $2$ which is the
  dimension of conformal field theory. In this case
  the scaling dimension with no spin 
is the sum of the holomorphic conformal dimension($\Delta$)
  and its antiholomorphic counterpart($\overline{\Delta}$). 
That is, $\Delta +\overline{\Delta} = 2\Delta < 2$. For
  example, the primary field $\Phi_{(1^n|2,1^{n-1})}^{(p)}$ is also
  a relevant field because its conformal dimension is less than $1$ due
  to (\ref{halfpert}). However, this relevant field is a `strongly'
  relevant field and so one cannot analyze perturbatively. On the other
  hand, perturbative analysis is possible for the `slightly'
  relevant field which has conformal dimension close to $1$.  }. 
More explicitly, one can compute the conformal dimension for the
relevant field (adjoint representation) from (\ref{Delta}) as follows:
\bea
\Delta_{(1^{n}|1,2, 1^{n-2})}^{(p)} & = &
\frac{(p-N+3)}{(p+1)} 
\simeq 1-\lambda, \qquad \lambda \equiv \frac{N}{k+N},
\label{pert}
\eea
where  we take the large $(N,k)$ 't Hooft 
limit with fixed 't Hooft coupling 
$\lambda$ defined as (\ref{limit}) of \cite{GG} in the last line of (\ref{pert})
here.
In the context of \cite{Fateev} where perturbation by an
appropriate operator leads to an IR fixed point described by the coset
$\frac{\widehat{SO}(N)_{k-1} \oplus
  \widehat{SO}(N)_1}{\widehat{SO}(N)_{k}}$(this can be obtained from
(\ref{coset}) by replacing $k$ with $k-1$), 
one can understand that the conformal dimension for an appropriate field
is given by the dual Coxeter number and the levels    
to be 
$\Delta_{(1^{n}|1,2, 1^{n-2})}^{(p)} =1-\frac{h^{\nu}}{k+1+h^{\nu}}=
1-\frac{N-2}{p+1}$ which is exactly the same as (\ref{pert}).
Of course, this description for the $A_{n-1}^{(p)}$ minimal model can be
analyzed and it can be seen that 
the behavior of (\ref{pert}) has features in common with the
conformal dimension of the adjoint representation of $A_{n-1}$.
It is not obvious how one can obtain 
the conformal dimensions from the coset model itself \cite{BG}. See
the papers \cite{ABL} or \cite{GG}  for the explicit formula.
From the quadratic Casimir $(N-2)$ for the adjoint representation of
$SO(N)$,
one can write down the conformal dimension as $1-\frac{(N-2)}{(N-2) +
k +1} = 1-\frac{N-2}{p+1}$ which is exactly the same as (\ref{pert}).  
That is, the first and second representations in the coset model (\ref{coset}) 
are trivial representations of $SO(N)$ while the diagonal representation is
the adjoint representation of $SO(N)$.
Here the quadratic Casimir is defined as
$\frac{1}{4}(\vec{n}^2-\vec{\rho}^2)$ for the representation $\vec{n}$
of $SO(N)$ and we will use this formula in the remaining parts of this
paper.

We noticed that the identity operator has a conformal
dimension of zero.
What is the lowest dimension operator, after the identity operator, in the singlet sector?
What happens if we take $(2,1^{n-1})$, which is a defining representation of 
$D_n$, as the $\alpha_{-}$ side as well as
the trivial $\alpha_{+}$ side $(1^n)$?
One computes the conformal dimension for the primary field 
$\Phi_{(1^n|2,1^{n-1})}^{(p)}$ 
exactly and takes the large
$(N,k)$ 't Hooft limit as before to obtain
\bea
\Delta_{(1^n|2, 1^{n-1})}^{(p)} & = & 
\frac{(p-N+2)}{2(p+1)} \simeq \frac{1}{2}(1-\lambda).
\label{halfpert}
\eea
This primary field is identified with the energy operator in
\cite{DNS}. Note that the factor $\frac{1}{2}$ comes from 
the $(1,1)$ component of the quadratic form matrix which is equal to
$2$ (see footnote \ref{quadform}),
together with the overall factor $\frac{1}{4}$ in the formula (\ref{Delta}).
Obviously at finite $(N,k)$, this expression is different from that 
of the fundamental representation of the $A_{(n-1)}^{(p)}$ minimal model.
However, they have common behavior in the large $(N,k)$ 't Hooft limit.
Furthermore, one can compute the conformal dimension
$\Delta_{(1^n|1^{n-1},2)}^{(p)}$ when the integer $2$ arises as the
last Dynkin label rather than the first label as in (\ref{halfpert}).
From the relation (\ref{expansion}), the constant piece looks like the
$(n,n)$-component of the quadratic form matrix, which is equal to $\frac{N}{4}$. 
This is rather different to the $A_{n-1}^{(p)}$ minimal
model where the corresponding dimension behaves as $\frac{N-1}{N}$.
From the quadratic Casimir $\frac{1}{2}(N-1)$ for the defining representation in
$SO(N)$,
one can write down the conformal dimension as
$\frac{1}{2}(N-1)[\frac{1}{(N-2)+1}-
\frac{1}{(N-2) + k +1}] = \frac{1}{2}-\frac{N-1}{2(p+1)}$ 
which is exactly the same as (\ref{halfpert}) where we used the
quadratic Casimir $(N-2)$ for the adjoint representation in the denominator.  
The first representation of (\ref{coset}) is a trivial representation
of $SO(N)$.

The operator product expansions of the fields 
$\Phi_{(\vec{n}|\vec{n}')}^{(p)}$ and $\Phi_{(\vec{m}|\vec{m}')}^{(p)}$
are, in general, linear combinations of $\Phi_{(\vec{s}|\vec{s}')}^{(p)}$ with the
appropriate structure constants of the operator algebra.
The selection rules of the operator algebra may be described by the
Clebsch-Gordan series for the product of the finite-dimensional
representations
of the Lie algebra $D_n$ with highest weights specified by the sets of
the numbers $(n_i,n_i')$ and $(m_i,m_i')$ corresponding to the
weight vectors.
Although the structure constants are determined by three-point
correlation functions through the Coulomb gas formalism, it is a rather
nontrivial task to find them explicitly.  
Luckily, the four necessary integrals from the 
Coulomb gas formalism have been computed and 
the structure constants are written in terms of these integrals.
Eventually, the fusion rules between the two primaries (adjoint and
defining representations) 
described by (\ref{pert}) and (\ref{halfpert})
can be summarized by \cite{DNS}
\bea
\Phi^{(p)}_{(1^n|2,1^{n-1})} \otimes \Phi^{(p)}_{(1^n|2,1^{n-1})} & = &
\Phi^{(p)}_{(1^n|1,2,1^{n-2})} +\cdots,
\nonu \\
\Phi^{(p)}_{(1^n|1,2,1^{n-2})} \otimes \Phi^{(p)}_{(1^n|1,2,1^{n-2})} & = &
\Phi^{(p)}_{(1^n|1,2,1^{n-2})} +\cdots,
\nonu \\
\Phi^{(p)}_{(1^n|1,2,1^{n-2})} \otimes \Phi^{(p)}_{(1^n|2,1^{n-1})} & = &
\Phi^{(p)}_{(1^n|2,1^{n-1})} +\cdots,
\label{fusion}
\eea
where we have ignored the identity operator and the 
terms on the right hand side that are irrelevant 
(in the context of RG analysis).
The structure constants appearing in the right hand side are obtained 
from the three-point correlation functions of the unperturbed
$D_n^{(p)}$ model \cite{DNS}.
When we look at the operator product expansion between 
$\Phi^{(p)}_{(1^n|2,1^{n-1})}(z)$ and
$\Phi^{(p)}_{(1^n|2,1^{n-1})}(w)$, 
there exists a factor 
$(z-w)^{-4\Delta_{(1^n|2, 1^{n-1})}^{(p)}+2\Delta_{(1^{n}|1,2, 
1^{n-2})}^{(p)}}$ in the right hand side of the first equation of (\ref{fusion}). 
Substituting the conformal dimensions 
(\ref{pert}) and (\ref{halfpert}) into this
exponent, gives the factor $(z-w)^{\frac{6}{p+1}}$ which 
goes to $1$ in the large
$p$ limit. Then, the normal ordered field product \cite{BS} (the constant
term in the operator product expansion) 
of $\Phi^{(p)}_{(1^n|2,1^{n-1})}(z)$ and
$\Phi^{(p)}_{(1^n|2,1^{n-1})}(z)$, denoted by 
$(\Phi^{(p)}_{(1^n|2,1^{n-1})} \Phi^{(p)}_{(1^n|2,1^{n-1})})(z)$, is
given by
$\Phi^{(p)}_{(1^n|1,2,1^{n-2})}(z)$ up to the structure constant which
is equal to $\sqrt{2}$ for large $N$, as follows
\bea
(\Phi^{(p)}_{(1^n|2,1^{n-1})} \Phi^{(p)}_{(1^n|2,1^{n-1})})(z) \simeq
\Phi^{(p)}_{(1^n|1,2,1^{n-2})}(z).
\label{product}
\eea
In other words, in the large $(N, k)$ 't Hooft limit, the conformal
dimension (\ref{pert}) of the perturbing primary field (adjoint representation) 
is equal to
twice 
the conformal dimension (\ref{halfpert}) of the primary field (defining representation). 
This is a new feature under the large $(N,k)$ 't Hooft limit. 
For the $A_{n-1}^{(p)}$ minimal model, the normal ordered field product
between the fundamental representation and the anti-fundamental representation
of $A_{n-1}$ is the (perturbing) adjoint representation of $A_{n-1}$ 
in the large $(N,k)$ 't Hooft limit \cite{GG}.

There exists a new critical point corresponding to the zero of
the $\beta$-function
at nonzero $g$ \cite{LF,DNS}. Due to the decrease of the $c$-function along the RG
flow,
this new critical point should correspond to the critical behavior of 
the $D_n^{(p')}$ model with $p' < p$ \cite{Zam86}. Note the $p$-dependence of the
central charge (\ref{centralcharge}).  How does one determine $p'$ in
the RG analysis? 
The central charge at this new critical point can be determined by
substituting $g^{\ast} = 4(n-1)\frac{\epsilon}{C} +{\cal
  O}(\epsilon^2)$ (which is the
solution of the $\beta$-function, where the $\epsilon$ here is the same as the one
in (\ref{expansion})) into the expression for 
the central charge $c_N(p)$ expanded in $g$ \cite{LF}, together with 
(\ref{centralcharge}) and (\ref{expansion}). It is found to be
\bea
c_N(p)^{\ast} = c_N(p) -\frac{64 (n-1)^3 \epsilon^3}{C^2}= c_N(p)
-\frac{4n(n-1)(2n-1)}{p^3} 
\simeq c_N(p-1),
\label{newc}
\eea
where 
$C$ is the structure constant appearing in front of 
$\Phi^{(p)}_{(1^n|1,2,1^{n-2})}(w)$ on the right hand side of
the operator product expansion (\ref{fusion}) between 
$\Phi^{(p)}_{(1^n|1,2,1^{n-2})}(z)$ and 
$\Phi^{(p)}_{(1^n|1,2,1^{n-2})}(w)$
and it is given by \cite{LF,DNS}
\bea
C_{(1^n|1,2,1^{n-2})(1^n|1,2,1^{n-2})}^{(1^n|1,2,1^{n-2})} = 
\frac{4(n-1)}{\sqrt{n(2n-1)}} +{\cal O}(\epsilon).
\label{structure}
\eea
The correction term in (\ref{newc}) comes from $-12 (n-1) \epsilon g^2 + 2 C
g^3 +\cdots$ at $g=g^{\ast}$.
The field theory, given by (\ref{modaction}) which has the UV behavior
described by the $D_n^{(p)}$ model, at $g > 0$, has also IR asymptotic behavior
that is 
described by the  $D_n^{(p-1)}$ model \footnote{
For $A_{n-1}^{(p)}$ minimal model, one can analyze similarly  and the 
central charge is $
c_N(p)^{\ast} = c_N(p) -\frac{8 n^3 \epsilon^3}{C^2}= c_N(p)
-\frac{2n(n^2-1)}{p^3} 
\simeq c_N(p-1)$
where 
$C$ is given by the result of the three-point correlation function at
leading order to be  
$
C_{(1^{n-1}|2,1^{n-3},2)(1^{n-1}|2,1^{n-3},2)}^{(1^{n-1}|2,1^{n-3},2)} 
= \frac{2n}{\sqrt{n^2-1}} +{\cal O}(\epsilon)$ \cite{LF}.
\label{ancase}}.
The equation (\ref{newc}) implies that 
the central charge at a nonzero fixed point agrees with that of the
$D_n^{(p-1)}$ model.
The perturbation of the coset theory by an appropriate operator 
$\Phi_{(1^n|1,2,1^{n-2})}^{(p)}$ 
changes $p$ into $(p-1)=p'$ where the difference $1$ is nothing but the
shift parameter (the level of the second spin 1 current
$E_{(2)}^{ab}(z)$ 
of $D_n$) of the coset (\ref{coset}). 
In the large $(N,k)$ 't Hooft limit, the RG flow changes the 't Hooft coupling, 
from $p$ to $p-1$(or $k$ to $k-1$),  as 
$
\delta \lambda =\frac{\lambda^2}{N}$
and this implies that
$
\delta c =- N \lambda \delta \lambda  =-\lambda^3
$ 
which can be seen from (\ref{newc}). We used the fact that
$c_N(\lambda) \simeq \frac{N}{2}(1-\lambda^2)$. 

In order to understand the IR behaviors of the primary fields, one
should consider the case where the $\alpha_{-}$ side is given by the trivial 
representation $(1^n)$.
That is, when the $\alpha_{+}$ side and $\alpha_{-}$ side for the weight
vector are
interchanged in (\ref{pert}) and (\ref{halfpert}),
one can compute the following dimensions for the defining representation
and an adjoint representation of $D_n$ exactly, as well as  
its large $(N,k)$ 't Hooft limit,  using the conformal dimension
formula
(\ref{Delta}) in order to see
how the primaries corresponding to (\ref{pert}) and (\ref{halfpert}) flow along the RG,
\bea
\Delta_{(2, 1^{n-1}|1^{n})}^{(p)} & = 
& \frac{(p+N-1)}{2p} \simeq \frac{1}{2}(1+\lambda),
\nonu \\
\Delta_{(1,2, 1^{n-2}|1^n)}^{(p)} & = &
\frac{(p+N-2)}{p} 
\simeq 1+\lambda.
\label{dimcon}
\eea
Note that the sum of (\ref{halfpert}) and the first equation of 
(\ref{dimcon}), for the defining representation, is equal to 1 under the
large $(N,k)$ 't Hooft limit. That is $\Delta_{(1^n|2, 1^{n-1})}^{(p)}
+\Delta_{(2, 1^{n-1}|1^{n})}^{(p)} \simeq 1$. Similarly, 
$ \Delta_{(1^{n}|1,2, 1^{n-2})}^{(p)} + \Delta_{(1,2,
  1^{n-2}|1^n)}^{(p)}
\simeq 2$.
The behavior of (\ref{dimcon}) in the large $(N,k)$ 't Hooft limit is the
same as those in the $A_{n-1}^{(p)}$ minimal model.
From the quadratic Casimir $\frac{1}{2}(N-1)$ for the defining
representation and quadratic Casimir $(N-2)$ for the adjoint
representation in
$SO(N)$,
one can write down the conformal dimensions, in the coset model directly, as
$\frac{1}{2}(N-1)[\frac{1}{(N-2)+k}+
\frac{1}{(N-2)  +1}] = \frac{1}{2}+\frac{N-1}{2p}$ and  
$1+\frac{(N-2)}{(N-2)+k}=1+\frac{N-2}{p}$. 
These coincide with (\ref{dimcon}) as we expected. 
In the former, the diagonal representation is a trivial representation
and in the latter, both the second and diagonal representations are
trivial ones. 
In all cases we use the formula for the quadratic Casimir that 
was given earlier.

The slope of the $\beta$-function at the fixed point \cite{Zam} provides 
the conformal dimension at the IR fixed point 
via $\frac{d \beta}{d g}|_{g^{\ast}}= - 2(n-1) \epsilon
+{\cal O}(\epsilon^2)$. 
Then the anomalous dimension for the relevant field (adjoint
representation) 
at the IR fixed point 
is given by
$
\Delta \simeq 1+\frac{2(n-1)}{p+1} \simeq 
1 +\frac{N}{p-1}$.
This is exactly the conformal dimension (\ref{dimcon}) of 
$\Phi_{(1,2, 1^{n-2}|1^n)}^{(p-1)}$ in $D_n^{(p-1)}$ minimal model
and therefore this leads to the flow 
\bea
\mbox{UV}: \Phi^{(p)}_{(1^n|1,2,1^{n-2})}(z) 
\longrightarrow  \mbox{IR}: \Phi^{(p-1)}_{(1,2,
  1^{n-2}|1^n)}(z).
\label{flow}
\eea
We also get a similar relation to (\ref{product}) in the IR region for
the $D_n^{(p-1)}$
model from the same analysis that was done in (\ref{product})
\bea
(\Phi^{(p-1)}_{(2,1^{n-1}|1^n)} \Phi^{(p-1)}_{(2,1^{n-1}|1^n)})(z)
\simeq \Phi^{(p-1)}_{(1,2,
  1^{n-2}|1^n)}(z).
\label{productother}
\eea
It is easy to see from (\ref{flow})  
how the flow of the primary field of the defining
representation of $D_n$ arises along the RG flow by realizing that 
the left hand side of (\ref{flow}) is given by the product of two
defining representations via (\ref{product}) and the right hand side
of (\ref{flow}) is given by the product of other defining
representations via (\ref{productother}). 

Alternatively, one can directly obtain the flow of the primary field of 
the defining representation. 
From (\ref{dimcon}) and (\ref{halfpert}), one also obtains
\bea
\Delta_{(2, 1^{n-1}|1^{n})}^{(p-1)}-\Delta_{(1^n|2, 1^{n-1})}^{(p)}  
= \left[\frac{1}{2} +\frac{N-1}{2(p-1)} \right] 
-\left[\frac{1}{2} -\frac{N-1}{2(p+1)}\right]
\simeq \lambda.
\label{Deltadiff}
\eea
On the other hand, the observation of Cardy and Ludwig \cite{LC} implies that 
the correction to the conformal dimension 
for small deviations from the new fixed point is  given by three
quantities:
two structure constants and the small parameter(which is related to
our minimal series index $p$) of the theory.
It is easy to check that 
\bea
\sqrt{\frac{2n-1}{n}} 
\left(\frac{4(n-1)}{\sqrt{n(2n-1)}} \right)^{-1}
 4(n-1) \epsilon = (2n-1) \epsilon \simeq \lambda,
\label{corr}
\eea
where we used the result of \cite{DNS} for the structure constant
appearing in the operator product expansion between 
$\Phi^{(p)}_{(1^n|2,1^{n-1})}(z)$ and 
$\Phi^{(p)}_{(1^n|2,1^{n-1})}(w)$ in the first equation of 
(\ref{fusion}) which is 
equal to $C_{(1^n|2,1^{n-1})(1^n|2,1^{n-1})}^{(1^n|1,2,1^{n-2})}
=\sqrt{\frac{2n-1}{n}} + {\cal O}(\epsilon)$ and another
structure constant given in (\ref{structure}). The last factor 
$ 4(n-1) \epsilon$ in (\ref{corr}) 
comes from the correction term of the central charge
(\ref{newc}). By comparing (\ref{Deltadiff}) with (\ref{corr}), 
in the IR, the field $\Phi_{(1^n|2, 1^{n-1})}^{(p)}$ of
the $D_n^{(p)}$ minimal model  is identified with
the field $\Phi_{(2, 1^{n-1}|1^{n})}^{(p-1)}$ of the $D_n^{(p-1)}$
minimal model and therefore one sees the flow 
\bea
\mbox{UV}:\Phi_{(1^n|2, 1^{n-1})}^{(p)} \longrightarrow 
\mbox{IR}:\Phi_{(2, 1^{n-1}|1^{n})}^{(p-1)}
\label{flowfun}
\eea
which is consistent with (\ref{product}) and (\ref{productother}) in the fact
that under the flow (\ref{flowfun}), the flow (\ref{flow}) is
satisfied as we mentioned before. 
For the $A_{n-1}^{(p)}$ minimal model, one can perform a similar analysis
and the computation of (\ref{Deltadiff}) gives
$\frac{N^2-1}{2N}(\frac{1}{p-1} +\frac{1}{p+1})
\simeq \lambda$. Although we do not know the structure constant
between the two primary fields of the
fundamental representations leading to the primary 
field of the adjoint representation(more precisely the coefficient of 
three-point function for these three fields), 
from the considerations of (\ref{Deltadiff}) and (\ref{flowfun}), one
concludes with the help of footnote \ref{ancase} 
that the large $(N,k)$ 't Hooft limit for this unknown
coefficient of the three-point function should be equal to $1$.
The analysis of the three-point function between   
the two primaries of the antifundamental representations and
the primary of the adjoint representation can be done similarly.  
%
The basic generating fields, from which we produce all the states in the
conformal field theory by taking the fusion products of them, 
are given by the following defining representations
\bea
\Phi^{(p)}_{(1^n|2,1^{n-1})} \qquad \mbox{and} \qquad 
\Phi^{(p)}_{(2,1^{n-1}|1^n)}.
\label{generating} 
\eea

From the operator product expansion in the first equation of
(\ref{fusion}),
one can think of the irrelevant fields having the next lowest conformal dimension.
From the Clebsh-Gordan coefficient between the two defining
representations of $SO(N)$, one obtains the conformal dimensions of the primary field 
$\Phi^{(p)}_{(1^n|3,1^{n-1})}$, where the $n_1'$ component in 
$\vec{n}'$ is greater than $1$.
Then one obtains the conformal dimension by using the formula (\ref{Delta}) and
moreover one can compute the conformal dimension for the other primary
field  $\Phi_{(3,1^{n-1}|1^n)}^{(p)}$ as follows:
\bea
\Delta_{(1^n|3, 1^{n-1})}^{(p)} & = & 
\frac{(2p-N+2)}{(p+1)} \simeq 2-\lambda,
\nonu \\
\Delta_{(3,1^{n-1}|1^n)}^{(p)} & = &
\frac{(2p+N)}{p} 
\simeq 2+\lambda.
\label{rel}
\eea
In this case, the quadratic Casimir for the $(3, 1^{n-1})$ representation
of $SO(N)$ is equal to
$N$. So the coefficient of the $N$-term in the first equation of (\ref{rel})
originates from $N[-\frac{1}{(N-2)+k+1}]=-\frac{N}{p+1}$ while the
coefficient of the $N$-term in the second equation comes from  
$N[\frac{1}{(N-2)+k}]=\frac{N}{p}$. 

How does one understand the primary field $\Phi^{(p)}_{(1^n|3,1^{n-1})}$
which has the conformal dimension given in the first relation of (\ref{rel})?
The one-loop contribution from the real scalar field in the bulk is
given by $Z_{\mbox{scal}}(h_{-}) = \prod_{l,l'=0}^{\infty} 
\frac{1}{(1-q^{h_{-}+l} \bar{q}^{h_{-}+l'})}$ 
where $h_{-}=\frac{1}{2}(1-\lambda)$ and  $q \equiv e^{2\pi i
  \tau}$. Here  $\tau$ is
the modular parameter which is the ratio of two complex 
periods of the lattice on a torus \cite{DMS}.
Expanding out the first few terms in $Z_{\mbox{scal}}(h_{-})$, one has 
a $q^{h_{-}} \bar{q}^{h_{-}}$ term, a $q^{2h_{-}} \bar{q}^{2h_{-}}$ term
and a $q^{2h_{-}+1} \bar{q}^{2h_{-}+1}$ term and so on. 
Since the conformal dimension for the adjoint representation 
is given by $\Delta_{(1^n|1,2,1^{n-2})}^{(p)}=2\Delta_{(1^n|2,
1^{n-1})}^{(p)}=2h_{-}$ in the large $(N,k)$ 't Hooft limit, 
eventually
the terms with an overall factor $q^{2h_{-}} \bar{q}^{2h_{-}}$
should correspond to the character for the adjoint 
representation $(1^n|1,2,1^{n-2})$ in the total partition function. 
Here we should add the contribution
$Z_{\mbox{hs}}$ (the explicit form will be given later) 
from
the gravitons of the higher spin fields. 
Similarly, the conformal dimension for the above irrelevant field
is given by $\Delta_{(1^n|3,1^{n-1})}^{(p)}=
2\Delta_{(1^n|2, 1^{n-1})}^{(p)} +1=2h_{-}+1$ in  the large $(N,k)$ 't
Hooft limit and 
the terms with an overall factor $q^{2h_{-}+1} \bar{q}^{2h_{-}+1}$
should correspond to the character for the 
representation $(1^n|3,1^{n-1})$  in the total partition function
which contains $Z_{\mbox{hs}}$.

Note that for the $A_{n-1}^{(p)}$ minimal model, the adjoint
representation appears in the fusion product of fundamental and
antifundamental representations and the fusion product of two
fundamental representations give other representations.
However, in the $D_n^{(p)}$ minimal model of this paper,
the adjoint representation arises from the fusion product of
two defining representations. 
The one-loop contribution from the other real scalar field in the bulk is
given by 
$Z_{\mbox{scal}}(h_{+}) = \prod_{l,l'=0}^{\infty} 
\frac{1}{(1-q^{h_{+}+l} \bar{q}^{h_{+}+l'})}$ 
where $h_{+}=\frac{1}{2}(1+\lambda)$.
Expanding out the first few terms, one obtains 
a $q^{h_{+}} \bar{q}^{h_{+}}$ term, a $q^{2h_{+}} \bar{q}^{2h_{+}}$ term
and a $q^{2h_{+}+1} \bar{q}^{2h_{+}+1}$ term. 
Since the conformal dimension for the adjoint representation 
is given by $\Delta_{(1,2,1^{n-2}|1^n)}^{(p)}=2\Delta_{(2,
1^{n-1}|1^n)}^{(p)}=2h_{+}$ in the large $(N,k)$ 't Hooft limit, 
the terms with an overall factor $q^{2h_{+}} \bar{q}^{2h_{+}}$
should correspond to the character for the 
adjoint representation $(1,2,1^{n-2}|1^n)$ in the total partition function. 
Similarly, the conformal dimension for the above irrelevant field
is $\Delta_{(3,1^{n-1}|1^n)}^{(p)}=
2\Delta_{(2, 1^{n-1}|1^n)}^{(p)} +1=2h_{+}+1$ in  the large $(N,k)$ 't
Hooft limit and 
the terms with an overall factor $q^{2h_{+}+1} \bar{q}^{2h_{+}+1}$
should correspond to the character for the 
representation $(3,1^{n-1}|1^n)$ in the total partition function where 
the contribution from $Z_{\mbox{hs}}$ should be added.

For the fusion product
$
\Phi^{(p)}_{(1^n|2,1^{n-1})} \otimes \Phi^{(p)}_{(2,1^{n-1}|1^n)}  = 
\Phi^{(p)}_{(2,1^{n-1}|2,1^{n-1})}
$ from
different types of combinations in (\ref{generating}), 
one can compute the conformal dimension for the 
primary field appearing in the right hand side 
and see that it is given by
$
\Delta_{(2,1^{n-1}|2,1^{n-1})}^{(p)}  = 
\frac{(N-1)}{2p(p+1)} 
\simeq \frac{\lambda^2}{2N}$.
This is consistent with the computation from the coset model
$\frac{1}{2}(N-1)[\frac{1}{(N-2)+k}-\frac{1}{(N-2)+k+1}]$ with  the
quadratic 
Casimir $\frac{1}{2}(N-1)$ for the $(2, 1^{n-1})$ representation
of $SO(N)$ as before. The second representation of the coset is a trivial one.
This is equal to the nonconstant piece on
the left hand side of fusion rule.
In other words, we have $\Delta_{(1^n|2, 1^{n-1})}^{(p)}
+\Delta_{(2, 1^{n-1}|1^{n})}^{(p)} = 1 +\frac{(N-1)}{2p(p+1)}$. 

What is the $AdS_3$ dual gravity theory of the two-dimensional coset
minimal model?
The primary field 
$\Phi^{(p)}_{(1^n|1,2,1^{n-2})}(z)$ is the normal ordered 
product of $(\Phi^{(p)}_{(1^n|2,1^{n-1})}
\Phi^{(p)}_{(1^n|2,1^{n-1})})(z)$ in (\ref{product}) and the
perturbation can be rewritten
\bea
 g \int d^2 x \,\, (\Phi^{(p)}_{(1^n|2,1^{n-1})}
\Phi^{(p)}_{(1^n|2,1^{n-1})})(x) =g \int d^2 x \,\, ({\cal O} {\cal O})(x),
\label{OO}
\eea
where the primary field ${\cal O}(z) \equiv \Phi^{(p)}_{(1^n|2,1^{n-1})}(z)$ has
holomorphic conformal dimension 
$\frac{1}{2}(1-\lambda)$ (\ref{halfpert}) in the large $(N,k)$ 't
Hooft limit. Its antiholomorphic
conformal dimension is also $\frac{1}{2}(1-\lambda)$. 
In the $AdS_3$ gravity theory side from the AdS/CFT correspondence
\cite{Witten01},  
the scalar field,
corresponding to ${\cal O}(z)$, with dimension
$\Delta_{-}$(which is the sum of holomorphic and antiholomorphic
conformal dimensions) is quantized in the $(-)$ quantization in the
UV (see also the relevant paper \cite{BSS}). 
In other words, the scalar field behaves as $\phi \sim
r^{1-\lambda}$ with an appropriate 
boundary condition where $r$ is a radial coordinate in
$AdS_3$ space. 
There exists an alternative choice for the quantization with
an irrelevant perturbation by an operator of dimension
$2-(1-\lambda)=(1+\lambda)$, 
where $\phi'$ behaves as $r^{1+\lambda}$,  but
this is not the case in (\ref{OO}). 
Along
the RG flow, this scalar field $\phi$ flows to the theory with $(+)$
quantization in the IR where it corresponds to an operator 
${\cal O}'(z) \equiv \Phi^{(p-1)}_{(2,1^{n-1}|1^n)}(z)$ with dimension
$\frac{1}{2}(1+\lambda)$ in the large $(N,k)$ 't
Hooft limit. The $({\cal O}{\cal O})(z)$ in (\ref{OO}) flows to 
an irrelevant operator of the form $({\cal O}'{\cal O}')(z)$. 
The two solutions for the mass formula of matter multiplet 
$M^2 = \Delta(\Delta-2)$
in higher spin theory are written as, by summing over holomorphic and
antiholomorphic parts, 
\bea
\Delta_{-} = \frac{1}{2} (1-\lambda) + \frac{1}{2} (1-\lambda) =
1-\lambda, \qquad
\Delta_{+} = \frac{1}{2} (1+\lambda) + \frac{1}{2} (1+\lambda) =
1+ \lambda.
\label{Deltas}
\eea
Therefore, the two real scalar fields $(\phi,\phi')$ in the $AdS_3$ gravity theory
with $M^2 = -(1-\lambda^2)$ where $\phi$ is in the $(-)$ quantization and 
$\phi'$ in the $(+)$ quantization match with the results for the
RG flow in the two-dimensional dual conformal field theories we
have described so far. 
Note that from (\ref{Deltas}) we have a relation $\Delta_{-}
+\Delta_{+} =2$ or $\Delta_{{\cal O}} +\Delta_{{\cal O}'}$=1 from
(\ref{halfpert})
and (\ref{dimcon}). See also the relevant work \cite{HK}  
for the changing of conformal dimension
$(1-\lambda)$ into $(1+\lambda)$ in the
different context of gravitational dressing (see also \cite{GK}).  
 
By following the procedure \cite{GGS} for the one-loop determinant in the heat
kernel techniques, one expects that 
the total one-loop determinant 
is given by the multiple product of each contribution for spin $s$.  
Then this can be interpreted using the boundary theory.
The vacuum character for the simply laced algebra with level $1$ is
given by \cite{BS,Bouwknegt} 
\bea
\chi = \frac{1}{\prod^{n}_{i=1} F_{e_i+1}(q) }, \qquad F_s(q)
\equiv \prod_{k=s}^{\infty} (1-q^k), \qquad q \equiv e^{2\pi i \tau}.
\label{char}
\eea
This is the vacuum character of 
type ${\cal W}(e_1+1, e_2+1, \cdots, e_{n}+1)$ algebra in the notation
of \cite{BS}.
For the $A_{n-1}^{(p)}$ minimal model, the algebra consists of 
a spin $2$ Virasoro generator and additional primary currents of 
spins $3, 4, \cdots, n(=N)$. 
Now let us apply the $SO(N)$ group to (\ref{char})
and realize that there exist $n$ exponents of $SO(N)$:
$e_1 =1, e_2=3, \cdots, e_{n-1}=2n-3$ and $e_{n}=n-1$. 
By taking into account the
antiholomorphic part,
the large $N$ limit of (\ref{char})  can be written as 
\bea
Z_{\mbox{hs}} = \lim_{N \to \infty}
\left(\prod_{m=2}^{\infty} 
\frac{1}{|1-q^m|^2}
\prod_{m=4}^{\infty} 
\frac{1}{|1-q^m|^2} \,\,\,
\cdots \,\,\,
\prod_{m=N-2}^{\infty} 
\frac{1}{|1-q^m|^2}
\prod_{m=\frac{N}{2}}^{\infty} 
\frac{1}{|1-q^m|^2} \right).
\label{hs}
\eea
This partition function from the $D_{n}^{(p)}$ minimal model 
conformal field theory should agree with that 
from the one-loop result in the higher spin bulk theory. 
Moreover, 
the higher spin theory we are interested in has two real scalar fields. 
The one-loop contributions from each scalar field can be obtained from 
\cite{GMY}. We also present the successive fusion products in the
context of the conformal field theory partition function.
The identifications,
\bea
Z_{\mbox{scal}}(h_{-}) = \prod_{l,l'=0}^{\infty} 
\frac{1}{(1-q^{h_{-}+l} \bar{q}^{h_{-}+l'})} 
\,\,\,\, & \leftrightarrow & 
\,\,\,\, \Phi^{(p)}_{(1^n|2,1^{n-1})} \otimes
 \cdots \otimes
\Phi^{(p)}_{(1^n|2,1^{n-1})}, 
\nonu \\
Z_{\mbox{scal}}(h_{+}) = \prod_{l,l'=0}^{\infty} 
\frac{1}{(1-q^{h_{+}+l} \bar{q}^{h_{+}+l'})} \,\,\,\, & \leftrightarrow &
\,\,\,\, 
\Phi^{(p)}_{(2,1^{n-1}|1^n)} \otimes
 \cdots 
\otimes
\Phi^{(p)}_{(2,1^{n-1}|1^n)}, 
\label{sca}
\eea
where 
$ h_{-} \equiv \frac{1}{2} \Delta_{-}$ and $h_{+} \equiv \frac{1}{2} \Delta_{+}$
imply that 
the left hand side of the first equation in 
(\ref{sca}) provides the contributions to the fusion product 
that contain the multiple copies of $\Phi^{(p)}_{(1^n|2,1^{n-1})}$ by
extending the simplest product  to the more general case. 
On the other hand, the left hand side of the second equation of (\ref{sca}) 
corresponds to the multiple copies of $\Phi^{(p)}_{(2,1^{n-1}|1^n)}$.
Then the total partition function can be written in terms of 
the partition functions in (\ref{hs}) and (\ref{sca}) as
\bea
Z_{\mbox{tot}}=
(q \bar{q})^{-\frac{c}{24}} Z_{\mbox{hs}} \, \, 
Z_{\mbox{scal}}(h_{-}) \, Z_{\mbox{scal}}(h_{+}).
\label{total}
\eea
In order to see the one-to-one correspondence precisely, the
computation for $Z_{\mbox{hs}}$ (\ref{hs}) should also be done  in the
bulk to see whether it really coincides with (\ref{hs}), which was obtained from the
computation in the boundary. Moreover, we described some
identifications in (\ref{sca}) but we did not show explicitly how the characters
in the boundary exactly match with $Z_{\mbox{scal}}(h_{\mp})$ (\ref{sca})
obtained from the bulk.     
According to \cite{Gtalks}, 
they have a conformal character \cite{LF,BS} and take the large
$(N,k)$
't Hooft limit. The branching function contains the character of
$U(\infty)$
and furthermore the scalar partition functions in (\ref{sca}) can be written in terms of
the characters of the 
representations of $U(\infty)$ \footnote{Just after
this paper was released in the arXiv, the two relevant papers 
\cite{GGHR} and \cite{GV} appeared in the arXiv. The former is the
published version of \cite{Gtalks} in which the partition
function of the $WA_{N-1}$ minimal model was obtained. 
The latter deals with the partition function of the
$WD_{\frac{N}{2}}$ minimal model. One of the main results of
\cite{GGHR} is as follows. Due to the fact that certain states become
null and decouple from correlation functions (and therefore have to be
removed from the spectrum), the careful limiting procedure shows that the resulting
states that survive exactly match the gravity prediction. The
simplest example is given by the fusion product
$\Phi^{(p)}_{(2,1^{n-1}|1^n)} \otimes \Phi^{(p)}_{(1^n|2,1^{n-1})}$
where both $\alpha_{+}$ and $\alpha_{-}$ are nontrivial. The conformal
dimensions are not additive. That is, $1=\Delta^{(p)}_{(2,1^{n-1}|1^n)}+
\Delta^{(p)}_{(1^n|2,1^{n-1})} \neq
\Delta^{(p)}_{(2,1^{n-1}|2,1^{n-1})}$.
However, their analysis shows that there exists a descendant state with
the conformal dimension $\Delta^{(p)}_{(2,1^{n-1}|2,1^{n-1})} = 
\Delta^{(p)}_{(2,1^{n-1}|1^n)}+
\Delta^{(p)}_{(1^n|2,1^{n-1})}=1$ in the conformal field theory
representation labeled by $(2,1^{n-1}|2,1^{n-1})$. This becomes the
generating state of the representation and the state $\psi$ 
and its descendants $\rho$ and $\xi$ in $(2.20)$ of
\cite{GGHR} match with the gravity results. The $\omega$ becomes null 
and the $\omega$ and its
descendants then decouple from the correlation functions. In this
computation, they considered the `strict' infinite $N$ limit where the
sum of the number of boxes and antiboxes in the Young tableau 
has maximum value in the
conformal field theory partition function.

What about the $WD_{\frac{N}{2}}$ minimal model case?
The above feature is related to the cross terms in the product of
$Z_{\mbox{scal}}(h_{+})$ and $Z_{\mbox{scal}}(h_{-})$ in (\ref{sca}).
According to the result of \cite{GV}, 
the character in the conformal field theory 
partition function consists of a linear combination of  
the Schur functions on the trivial representation $(1^n)$, 
on  the adjoint representation $(1,2,1^{n-2})$ and 
on the representation $(3,1^{n-1})$. It turns out that the
states corresponding to the Schur function on the trivial
representation, which  are the states generated from $\omega$, become null and 
decouple from the correlation function. Of course, 
the states from $\psi$ and its descendants corresponding to 
the Schur functions on  the adjoint representation $(1,2,1^{n-2})$ and 
on the representation $(3,1^{n-1})$ match the gravity prediction.
The total conformal field theory partition function agrees with the
bulk partition function.  }.
From this observation for the $A_{n-1}^{(p)}$ minimal model 
it may be that one can also write down the scalar partition
functions in terms of a sum over the characters of representations of 
$SO(\infty)$ (or its more general group $O(\infty)$). 
In order to understand this clearly, it is useful
to look at the expansion of characters developed in \cite{Balantekin,Balantekin1}.   
The sum over the Weyl group elements and the sum over 
the lattice (generated by the simple roots of the Lie algebra $D_n$) 
in the character formula \cite{LF} should be related to   
the sum over the characters of representations of 
$SO(N)$ in the large $N$ limit.

\section{ The large $(N, k)$ limit of coset minimal $W B_n^{(p)}$ model}

Let us consider the same `diagonal' coset model (\ref{coset})
where a rank $n$ for the non-simply laced algebra $B_n=SO(2n+1)$ has a relation 
\bea
N \equiv 2n+1.
\label{rank1}
\eea
The central charge is given by (\ref{centralcharge}) with
the minimal model index (\ref{pcondition}). 
For $N=5$, some
coset theories with different choices of levels are described in \cite{Ahn91,Ahn92,DE06}.
The primary operators of the minimal model are represented by 
the vertex operators that can be associated with the weight lattice of 
$B_n$(or $B_{\frac{N-1}{2}}$ via (\ref{rank1})) \cite{LF}. 
The Coulomb gas formula for the conformal dimension of the 
primary operator $\Phi_{(\vec{n}|\vec{n}')}^{(p)}$ in the Neveu-Schwarz
sector
where $(n_n-n_n')$ is even
can be summarized by \cite{LF,DE,DJR}
\bea
\Delta_{(\vec{n}|\vec{n}')}^{(p)} =\frac{1}{2p(p+1)}
\left[((p+1)\vec{n}-p\vec{n}')^2-\vec{\rho}^2 \right], \qquad 
\vec{\rho}^2 = \frac{1}{12} n (2n-1)(2n+1).
\label{Delta1}
\eea
For the Ramond-Ramond sector where $(n_n -n_n')$ is odd, there is an extra $\frac{1}{16}$
factor in the above dimension formula.
More explicitly, one can compute the conformal dimensions for the
lowest dimensional field and the relevant field from (\ref{Delta1})
respectively 
as follows:
\bea
\Delta_{(1^n|2, 1^{n-1})}^{(p)} & = & 
\frac{(p-N+2)}{2(p+1)} \simeq \frac{1}{2}(1-\lambda),
\nonu \\
\Delta_{(1^{n}|1,2, 1^{n-2})}^{(p)} & = &
\frac{(p-N+3)}{(p+1)} 
\simeq 1-\lambda.
\label{basic}
\eea
Although the quadratic form matrix for the $B_n$ group \footnote{For
convenience, we present the elements here: $\vec{w}_i\cdot\vec{w}_j= i$ for
$i\leq j < n$, $\vec{w}_i\cdot\vec{w}_n= \frac{i}{2}$ for $i < n$ and 
$\vec{w}_n\cdot\vec{w}_n= \frac{n}{4}$. }
is different from that
of the $D_n$ group(in footnote \ref{quadform}) 
and the expression for the conformal dimension (\ref{Delta1})
looks different from that (\ref{Delta}) of $D_n$, the expressions 
(\ref{pert}) and (\ref{halfpert}) at finite $N$ and $k$ are
coincident with (\ref{basic}). 

In the original paper \cite{LF}, the RG analysis was described for
$A_{n-1}^{(p)}$
and $D_n^{(p)}$ models only 
but the $B_n^{(p)}$ model can also be analyzed in a similar way.
For example, the perturbed action is the same as in (\ref{modaction}).
The normal ordered product (\ref{product}) holds by taking the large
$(N,k)$
't Hooft limit.
Then the central charge at the new critical point can be determined by
substituting $g^{\ast} = 2(2n-1)\frac{\epsilon}{C}+ {\cal
  O}(\epsilon^2)$ 
into the expression of 
the central charge $c_N(p)$ expanded in $g$
\bea
c_N(p)^{\ast} = c_N(p) -\frac{8 (2n-1)^3 \epsilon^3}{C^2}= c_N(p)
-\frac{2n(2n-1)(2n+1)}{p^3} 
\simeq c_N(p-1),
\label{ccorr}
\eea
where 
$C$ is the structure constant corresponding to (\ref{structure}) for
the $B_n^{(p)}$
minimal model and  is given by \cite{DE}, via the three-point
function in the Coulomb gas representation(that is, the fusion
constant and the normalization of the vertex operator), to be
\bea
C_{(1^n|1,2,1^{n-2})(1^n|1,2,1^{n-2})}^{(1^n|1,2,1^{n-2})} 
= \frac{2(2n-1)}{\sqrt{n(2n+1)}} +{\cal O}(\epsilon).
\label{str}
\eea
Of course, the motivation of \cite{DE} is to describe the RG flows for
the second parafermion theory which will be described in next section 
but, as a  by-product, they also  found this
structure constant through the Coulomb gas representation with a
three-point function. 
Similar analysis gives the flow (\ref{flow}) for the $B_n^{(p)}$ minimal
model 
under the RG flow with (\ref{productother}).
One obtains the following conformal dimensions, corresponding to
(\ref{basic}) but with the $\alpha_{+}$ side and the 
$\alpha_{-}$ side interchanged,
which allow us to understand how the primary fields transform under the RG flow, 
\bea
\Delta_{(2, 1^{n-1}|1^{n})}^{(p)} & = 
& \frac{(p+N-1)}{2p} \simeq \frac{1}{2}(1+\lambda),
\nonu \\
\Delta_{(1,2, 1^{n-2}|1^n)}^{(p)} & = &
\frac{(p+N-2)}{p} 
\simeq 1+\lambda.
\label{basic1}
\eea
These match the conformal dimensions (\ref{dimcon})
for the $D_n^{(p)}$ model meaning that the two models show the 
same behavior.

The correction of the conformal dimension 
for a small deviation from the new fixed point can be
written as
\bea
\frac{2n}{\sqrt{n(2n+1)}} \left(\frac{2(2n-1)}{\sqrt{n(2n+1)}}\right)^{-1}
 2(2n-1) \epsilon = 2n \epsilon \simeq \lambda,
\label{difference}
\eea
where the structure constant
$C_{(1^n|2,1^{n-1})(1^n|2,1^{n-1})}^{(1^n|1,2,1^{n-2})}=
\frac{2n}{\sqrt{n(2n+1)}} +{\cal O}(\epsilon)$ was found in
\cite{DE} and the structure constant (\ref{str}) is used.
The fusion rules (\ref{fusion}) are also valid for this case.
In addition, the factor $ 2(2n-1) \epsilon$ is consistent with the correction
term for the central charge in (\ref{ccorr}).
On the other hand, there is a   difference between the conformal
dimensions, which can be computed from (\ref{basic}) and
(\ref{basic1}) to be
\bea
\Delta_{(2, 1^{n-1}|1^{n})}^{(p-1)}-\Delta_{(1^n|2, 1^{n-1})}^{(p)}  
= \left[\frac{1}{2} +\frac{N-1}{2(p-1)} \right] -\left[\frac{1}{2} -\frac{N-1}{2(p+1)}
\right] 
\simeq \lambda.
\label{diff1}
\eea
By comparing (\ref{difference}) with (\ref{diff1}), 
in the IR, the field $\Phi_{(1^n|2, 1^{n-1})}^{(p)}$ of the $B_n^{(p)}$ 
minimal model is identified with
the field $\Phi_{(2, 1^{n-1}|1^{n})}^{(p-1)}$ of the $B_n^{(p-1)}$
minimal model.
The relation (\ref{flowfun}) also holds  for the $B_n^{(p)}$ minimal model.

According to \cite{BS}, the vacuum character for $B_n$
with level $1$ has an extra contribution from
the fermionic $(n+\frac{1}{2})$ dimensional field  projected onto the
$Z_2$
even sector. Odd $Z_2$ parity is assigned to the currents of 
half odd integer spin and even $Z_2$ parity is assigned to
the integer spin currents \cite{BS}. 
The singlet algebra is the bosonic projection of the type 
${\cal W}(2, 4, \cdots, 2n=N-1, n+\frac{1}{2}=\frac{N}{2})$. 
Then the large 
$N$ limit of the partition function for the higher spin with field
contents (\ref{wb}) is written as
\bea
\lim_{N \to \infty}
\left(\prod_{m=2}^{\infty} 
\frac{1}{|1-q^m|^2}
\,\,\,
\cdots \,\,\,
\prod_{m=N-1}^{\infty} 
\frac{1}{|1-q^m|^2} \times 
|\frac{1}{2} [\prod_{m=\frac{N}{2}}^{\infty} (1+q^{m+\frac{1}{2}}) +
\prod_{m=\frac{N}{2}}^{\infty} (1-q^{m+\frac{1}{2}})]|^2 \right),
\label{bhs}
\eea
where the last term  in (\ref{bhs}) is the vacuum character of the above
fermion field projected onto the $Z_2$ even sector. 
This is very similar to the bosonic projection of the ${\cal N}=1$
superconformal algebra, which can be realized as the $WB_1$ minimal model
because the field contents from (\ref{wb}) are given by a spin 2
Virasoro generator and
spin $\frac{3}{2}$ superpartner, 
of type ${\cal W}(2, 4, 6)$ \cite{Bouwknegt}. See also \cite{Watts}
for the coset currents of spin $(n+\frac{1}{2})$ and representation theory.
Finally, one obtains the total partition function (\ref{total}) where 
the higher spin part $Z_{\mbox{hs}}$ is given by (\ref{bhs}) for
the $B_n^{(p)}$ minimal model.

\section{Conclusions and outlook }

We described the dualities between the large $(N,k)$ 't Hooft limits of the 
$WD_n^{(p)}$ and $WB_n^{(p)}$ coset minimal models and 
the higher spin theory on $AdS_3$ where two massive real scalars
are added to the massless higher spin fields. We explained 
this duality by showing that the RG flows of the two-dimensional
conformal field theories agree with the gravity analysis from 
the AdS/CFT correspondence. 

So far, the level of the second group in the coset model is $1$.
What happens if the `shift parameter' is greater than 1?
We present two examples.
The first example, from a series of unitary conformal field theories, is 
the second parafermion theory  by Fateev and Zamolodchikov \cite{FZ1985}.
Note that the first parafermion theory is a single conformal field
theory for given $N$. The diagonal coset model, denoted by $Z_N^{(2)}(p)$, for $N\geq 5$ 
is characterized by \cite{GS,BG}
\bea
\frac{\widehat{SO}(N)_k \oplus \widehat{SO}(N)_2}{\widehat{SO}(N)_{k+2}}.
\label{paracoset}
\eea
The central charge for (\ref{paracoset}) is given by \cite{FZ1985}
as
\bea
c_N(p)  =
(N-1) \left[ 1- \frac{N(N-2)}{p(p+2)} \right] \leq (N-1), 
\qquad p\equiv k+N-2 \geq N-1,
\label{central}
\eea
which can be seen 
by realizing that the correct level for the second group 
in this case is $2$ rather than $1$.
In the large $(N,k)$ 't Hooft limit, this reduces to 
$c_N(p) \simeq N (1-\lambda^2)$ which is twice that of previous examples.
For $N=2$, the theory is given by $c=1$ free boson theory. 
For $N=3$ parafermion theory, 
developed in \cite{FZ}, it is known that the coset is given by
$\frac{\widehat{SU}(2)_{2k} 
\oplus \widehat{SU}(2)_4}{\widehat{SU}(2)_{2k+4}}$ where
$\widehat{SO}(3)_k$ is identified with $\widehat{SU}(2)_{2k}$.
The two slightly relevant perturbations on this coset model are
described in \cite{CSS,CPSS} and there exists only a single IR fixed
point denoted by $Z_3^{(2)}(p-4)$. 
For $N=4$, the parafermionic algebra factorizes into a direct product
of two ${\cal N}=1$ superconformal algebras. 
The slightly relevant perturbation on a single ${\cal N}=1$
superconformal algebra has been discussed in \cite{KIKKMO}.

According to the observation of Dotsenko and Estienne 
\cite{DE}, the two slightly relevant
fields (for odd $N\geq 7$ and for $N=5$, they also presented the
corresponding quantities),
can be obtained from the product of $WB_n$ primaries
$\Phi_{(\vec{n}|\vec{n}')}^{(p)}$ 
by decomposing the coset (\ref{paracoset}) into 
several simpler cosets
as follows
\bea
S^{(p)}_{(1^n|3,1^{n-1})} & = & \Phi^{(p)}_{(1^n|2,1^{n-1})} \otimes 
\Phi^{(p+1)}_{(2,1^{n-1}|3,1^{n-1})},
\nonu \\
A^{(p)}_{(1^n|1,2,1^{n-2})} & = & \frac{1}{\sqrt{2}} \left[
\Phi^{(p)}_{(1^n|1^n)} \otimes 
\Phi^{(p+1)}_{(1^n|1,2,1^{n-2})} + \Phi^{(p)}_{(1^n|1,2,1^{n-2})} \otimes 
\Phi^{(p+1)}_{(1,2,1^{n-2}|1,2,1^{n-2})} \right].
\label{SA}
\eea
These two fields appear in the following perturbed action
\bea
S^{(p)} = S_0^{(p)} + g \int d^2 x \,\, S^{(p)}_{(1^n|3,1^{n-1})}(x)
+ h \int d^2 x \,\, A^{(p)}_{(1^n|1,2,1^{n-2})}(x).
\label{Saction}
\eea
It is straightforward to compute the conformal dimensions for the
fields(Neveu-Schwarz sector) in (\ref{SA}) via (\ref{Delta1})
\bea
\Delta_{(1^n|2, 1^{n-1})}^{(p)} & = & 
\frac{(p-N+2)}{2(p+1)} \simeq \frac{1}{2}(1-\lambda),
\nonu \\
\Delta_{(2,1^{n-1}|3, 1^{n-1})}^{(p+1)} & = &
\frac{p(p-N+2)}{2(p+1)(p+2)} 
\simeq \frac{1}{2}(1-\lambda),
\nonu \\
\Delta_{(1^{n}|1^n)}^{(p)} & = & 0,
\nonu \\
\Delta_{(1^{n}|1,2, 1^{n-2})}^{(p+1)} & = &
\frac{(p-N+4)}{(p+2)} 
\simeq 1-\lambda,
\nonu \\
\Delta_{(1^{n}|1,2, 1^{n-2})}^{(p)} & = &
\frac{(p-N+3)}{(p+1)} 
\simeq 1-\lambda,
\nonu \\
\Delta_{(1,2,1^{n-2}|1,2, 1^{n-2})}^{(p+1)} & = &
\frac{(N-2)}{(p+1)(p+2)} 
\simeq \frac{\lambda^2}{N} \simeq 0.
\label{confdim}
\eea
As expected, the conformal dimensions for $ S^{(p)}_{(1^n|3,1^{n-1})}$
and 
$ A^{(p)}_{(1^n|1,2,1^{n-2})}$ in (\ref{Saction}), in the large
$(N,k)$ 't Hooft limit, can be read off from (\ref{confdim}) and they
become
$1-\lambda$. 
The exact expression for the conformal dimension of
$A^{(p)}_{(1^n|1,2,1^{n-2})}$
is $1-\frac{h^{\nu}}{k+2 +h^{\nu}} 
=1-\frac{N-2}{p+2}$ which can be seen from the result of \cite{Fateev}
where there exists only a single relevant field.
The first and second representations of (\ref{paracoset})
are trivial representations of $SO(N)$.
Moreover, the conformal dimension of  $ S^{(p)}_{(1^n|3,1^{n-1})}$
is given by $ 1-\frac{N}{p+2}$.
For large $p$, they have the same conformal dimension.
There exist two kinds of fixed points for nonzero $h$, which can be
seen
by analyzing the
RG flow from (\ref{Saction}). 
Dotsenko and Estienne \cite{DE} claim that
for the first kind of fixed point, the IR theory is described by
$Z^{(2)}_{N}(p-2)$
parafermion theory while 
for the second kind of fixed point, the IR theory
is given by $Z^{(2)}_{N}(p-1)$ parafermion theory.
The presence of $ S^{(p)}_{(1^n|3,1^{n-1})}$ in the perturbed action
(\ref{Saction}) provides the latter critical fixed point. 
With $ A^{(p)}_{(1^n|1,2,1^{n-2})}$ only, the former fixed point occurs. 

The deviation of the central charge from the two fixed points
can be computed from (\ref{central}) to be
\bea
\delta c = c_N(p-l)-c_N(p) \simeq - 2 l \lambda^3, \qquad l=1, 2.
\label{centralvar}
\eea
%
This can be seen 
by taking the variation 
$\delta c = -2N \lambda
\delta \lambda$ with $\delta k = -l$(from $k-l$ to $k$)
in the relation $c_N(\lambda) \simeq N (1-\lambda^2)$.
For $l=1$ in (\ref{centralvar}), 
the IR theory is given by  $Z^{(2)}_{N}(p-1)$ parafermion
theory
and for $l=2$, the IR theory is  $Z^{(2)}_{N}(p-2)$ parafermion theory.
One should also see this behavior (\ref{centralvar}) in the bulk. 
How do the adjoint primary fields (or their $WB_n$ products) 
flow under the RG flows? Although the particular primary field 
$\Phi_{(\vec{n}|\vec{n})}^{(p)}$ flows to
$\Phi_{(\vec{n}|\vec{n})}^{(p-l)}$
where $l=1, 2$ under the RG flow \cite{DE}, it is not known in general how the
other primaries flow. 
It is an open problem to find the gravity duals of the
above generalized conformal field theories.
For even $N(\geq 6)$, a similar construction is given in
\cite{Estienne}. See also \cite{DJS} for the details.
In
this case, the constructions (\ref{SA}) and (\ref{confdim}) are based
on the $WD_n^{(p)}$ primaries with the conformal dimension formula (\ref{Delta}).
It turns out that the conformal dimensions for $WD_n^{(p)}$ primaries
are the same as the ones in (\ref{confdim}).

Let us discuss the second example where the shift parameter is greater
than $1$.
Although the original motivation of \cite{GG} is to search for the
nontrivial example of 
nonsupersymmetric AdS/CFT correspondence, it is an interesting problem to
find a supersymmetric version of the proposal of \cite{GG}. 
For example, let us consider the diagonal coset model
\bea
\frac{\widehat{SU}(N)_3 \oplus \widehat{SU}(N)_k}{\widehat{SU}(N)_{k+3}}.
\label{Co}
\eea
The level $3$ is crucial for the construction of fermionic currents in
order to supersymmetrize the theory.
The central charge of (\ref{Co}) can be computed from the dual Coxeter
number and the dimension of the $SU(N)$ group and is written as  
\bea
c_N(p) & = & (N^2-1) \left[ \frac{3}{3+N} + 
\frac{k}{k+N} - \frac{k+3}{k+3+N} \right]
\nonu \\
&=&
\frac{3(N^2-1)}{N+3} \left[ 1- \frac{N(N+3)}{p(p+3)} \right] \leq \frac{3(N^2-1)}{N+3},
\qquad p \equiv k+N, k=1, 2, \cdots,
\label{c}
\eea
by considering the right levels.
In the large $(N,k)$ 't Hooft limit, this reduces to 
$c_N(p) \simeq 3N (1-\lambda^2)$. Again the factor $3$ comes from the 
level of the first group.
For $N=3$, the coset constructions and minimal series
are found in \cite{ASS}.
The spin $\frac{3}{2}$ fermionic superpartner of $\widetilde{T}(z)$, denoted by
$\widetilde{G}(z)$,
can be constructed as in \cite{GKO86} 
and the spin $3$ coset field $\widetilde{W}(z)$ can be determined by the
requirements \cite{BBSS} that it should be a primary field of dimension 3 with
respect to $\widetilde{T}(z)$ and the coefficient of the identity in
the operator product expansion $\widetilde{W}(z) \widetilde{W}(w)$
should be $\frac{c}{3}$ with (\ref{c}). 
Now one can compute the operator product expansion between 
$\widetilde{G}(z)$ and $\widetilde{W}(w)$ and it turns out
that the spin $\frac{5}{2}$ coset field $\widetilde{U}(z)$ is \cite{ASS,HR}
\bea
\widetilde{U}(z) = d_{abc} \left[ \frac{10\lambda^2}{(1-\lambda)(2-\lambda)}
    \psi_{(1)}^a 
V_{(2)}^b V_{(2)}^c(z) -\frac{5\lambda}{(1-\lambda)} \psi^a_{(1)}
V_{(1)}^b V_{(2)}^c(z) +\psi_{(1)}^a V_{(1)}^b 
V_{(1)}^c(z) \right],
\label{U}
\eea
where $\psi^a(z)$ is a free fermion field of dimension $\frac{1}{2}$
with $a=1, 2, \cdots, N^2-1$ and $V_{(1)}^a(z)$ is a spin $1$ current
that can be written in terms of free fermions as $V_{(1)}^a(z) =
f^{abc} (\psi_{(1)}^b \psi_{(1)}^c)(z)$ up to an overall $N$-dependent
constant with level $3$. 
Similarly, $V_{(2)}(z)$ is a spin $1$ current with level $k$.
Here the $d_{abc}$ symbol in (\ref{U}) 
is the symmetric traceless invariant tensor of rank $3$ for $SU(N)$.

Contrary to the description for 
the spin $3$ primary field $\widetilde{W}(z)$ \cite{GH}, for the above 
spin $\frac{5}{2}$ primary field, there is no vanishing term when we
take the large $(N,k)$ 't Hooft limit. This is due to the fact that, by
construction, there are no such terms coming from only the second
group with subscript $(2)$ and 
moreover there exists an overall factor $\psi_{(1)}^a(z)$ in (\ref{U}).   
Since the eigenvalues of the spin $3$ mode of the coset algebra corresponding to
$\widetilde{W}(z)$ in the large $(N,k)$ 't Hooft limit coincide with 
the eigenvalues of the zero mode of higher spin $3$ in the wedge algebra, 
one should expect that the above extended currents should preserve
the higher spin wedge algebra.  
The supersymmetric extension of \cite{PRS} appears in the work of \cite{BVW,BVW1}.
In the Neveu-Schwarz sector, there is a finite $OSp(1,2)$ subalgebra
generated by the $sl(2)$ generators $L_0, L_{\pm}$ for the Virasoro
generator and the mode $G_{\pm \frac{1}{2}}$
of its superpartner. 
It would be interesting to see how the supersymmetric higher spin
algebra  \cite{BVW,BVW1} is realized in the coset model (\ref{Co}) or
other unitary coset minimal models. 

Other possible supersymmetric versions of \cite{GG} can be studied by
using the quantum Drinfeld-Sokolov 
construction of the affine Lie superalgebra $\widehat{SU}(n+1,n)$ 
that provides the ${\cal N}=2$ super $W_n$ algebras \cite{BS}.

\vspace{.7cm}

\centerline{\bf Acknowledgments}

CA would like to thank  
the following people for discussions or correspondence on the
following topics:
V.S. Dotsenko on the RG flow, B. Estienne on
the correlation function, V.A. Fateev on the structure constant,
R. Gopakumar on the fusion rule and higher spin theory, 
T. Nishioka on $AdS_3$ gravity theory, T. Hartman on the zero mode of
spin $4$ field, I.R. Klebanov on double-trace deformation, 
and R. Volpato on the branching functions. 
In particular, CA thanks B. Estienne for discussions on the conformal
field theories.  
This work was supported by the Mid-career Researcher Program through
the National Research Foundation of Korea (NRF) grant 
funded by the Korean government (MEST) (No. 2009-0084601).
CA acknowledges warm hospitality and partial support from 
the Department of Physics, Princeton University.

\newpage





\end{document}